# Electric energies of a charged sphere surrounded by electrolyte

István P. Sugár

Neurology Department, Icahn School of Medicine at Mount Sinai,

New York, NY 10029

## Abstract

By using the generalized version of Newton's Shell Theorem [6] analytical equations are derived to calculate the electric energy of a charged sphere and the electrolyte's field energy inside and around the sphere. These electric energies are calculated as a function of the electrolyte's ion concentration. The work needed to build up the charged sphere, $E_{CC}$ (i.e. the total charge-charge interaction energy) decreases with increasing ion concentration of the electrolyte because of the electrolyte ions' screening effect on the charge-charge interaction. The energy needed to build up the charged sphere appears as sum of the electrolyte's field energy and the polarization energy of the electrolyte ions. At zero ion concentration the electrolyte's field energy is equal with $E_{CC}$ while the polarization energy is zero. At high ion concentrations ($C > 10[mol \cdot m^{-3}]$) 50% of the charge-charge interaction energy appears as the polarization energy of ions, 25% as the electrolyte's field energy inside the sphere and 25% as the electrolyte's field energy around the sphere.

**Keywords:** Debye length; screened potential; electrolyte's field energy; charge-charge interaction energy

## 1. Introduction

The head groups of membrane lipids have either single charge (e.g. tetraether lipids [1,2]) or electric dipole (e.g. phospholipids [1,3]). Theoretical models of lipid membranes usually focus on short range (Van der Waals) lateral interactions between nearest neighbor lipids and ignore the long range charge-charge interactions [3,4]. This is because in the case of long range interactions one has to consider the entire system rather than the interactions between the nearest neighbor lipids. In order to get closer to the solution of this problem recently we developed a generalized version of Newton's Shell Theorem [5,6]. According to the Generalized Shell Theorem the potential around a charged sphere of radius $R$ is (see Eq.9 in ref.6)

$$V(Z, Q) = \frac{k_e \cdot Q \cdot \lambda_D}{\varepsilon \cdot Z \cdot R} \cdot e^{-\frac{Z}{\lambda_D}} \cdot sinh\left(\frac{R}{\lambda_D}\right) \qquad \text{at } Z > R \qquad (1)$$

where $Z$ is the distance from the center of the charged sphere and the potential at $Z \leq R$ is (see Eq.10 in ref.6)

$$V(Z,Q) = \frac{k_e \cdot Q \cdot \lambda_D}{\varepsilon \cdot Z \cdot R} \cdot e^{-\frac{R}{\lambda_D}} \cdot sinh\left(\frac{Z}{\lambda_D}\right) \tag{2}$$

where $Q = 4R^2\pi\rho$ is the total charge of the sphere and $\rho$ is the surface charge density, $\lambda_D$ is the Debye length in the electrolyte that is inside and around the charged sphere, $k_e(= 9 \cdot 10^9 Nm^2 C^{-2})$ is the Coulomb's constant and $\varepsilon$ is the relative static permittivity of the electrolyte. In ref.6 we also calculated the electric potential of two concentric charged spheres surrounded by electrolyte, and the membrane potential of a charged lipid vesicle surrounded by electrolyte with high ion concentration. At any electrolyte concentration one can calculate the electric potential of the charged lipid vesicle by numerical integration (see ref. 6).

In this paper we consider a single charged sphere (as in ref.6) and calculate the charge-charge interaction energies and the electrolyte's field energy inside and around the charged sphere. By using the potential $V(Z,Q)$ of the charged sphere (Eqs.1,2) the electrolyte's field energy can be calculated in these two regions by the following equation [7]:

$$E_{region} = \frac{\varepsilon \cdot \varepsilon_0}{2} \int \underline{E} \cdot \underline{E} dv = \frac{\varepsilon \cdot \varepsilon_0}{2} \int \left[\frac{dV(Z,Q)}{dZ}\right]^2 4\pi Z^2 \cdot dZ \tag{3}$$

where $\varepsilon_0 (= [4\pi k_e]^{-1})$ is the vacuum permittivity. Since our considered system is central symmetric the direction of the electric field vector, $\underline{E}$ is the radial direction and thus $|\underline{E}| = \frac{dV}{dZ}$ and at each region the integration is taken from the lower to the upper radius of the region.

## 2. Model

To build a charged sphere it requires energy to overcome the repulsive forces between the charges of the sphere. According to the Generalized Shell Theorem, Eqs.1,2, the potential on the surface of the charged sphere of radius $R$ with charge $q$ is:

$$V(R,q) = \frac{k_e \cdot q \cdot \lambda_D}{\varepsilon \cdot R^2} \cdot e^{-\frac{R}{\lambda_D}} \cdot sinh\left(\frac{R}{\lambda_D}\right) \tag{4}$$

The charge-charge interaction energy needed to build a sphere of total charge $Q$ is:

$$E_{CC} = \int_0^Q V(R,q)dq = \frac{k_e \cdot Q^2 \cdot \lambda_D}{2 \cdot \varepsilon \cdot R^2} \cdot e^{-\frac{R}{\lambda_D}} \cdot sinh\left(\frac{R}{\lambda_D}\right) \tag{5}$$

The energy of the electric field inside the charged sphere, $Z \leq R$, is calculated by using Eqs.2,3 (see Appendix 1):

$$E_{in} = \frac{\varepsilon \cdot \varepsilon_0}{2} \int_0^R \left[\frac{dV(Z,Q)}{dZ}\right]^2 4\pi Z^2 dZ =$$

$$\frac{k_e Q^2 \lambda_D^2}{2\varepsilon R^2} e^{-\frac{2R}{\lambda_D}} \left\{\frac{1}{2\lambda_D} \sinh\left(\frac{R}{\lambda_D}\right) \cosh\left(\frac{R}{\lambda_D}\right) + \frac{R}{2\lambda_D^2} - \frac{1}{R}\sinh^2\left(\frac{R}{\lambda_D}\right)\right\} \quad (6)$$

The energy of the electric field around the charged sphere, $Z > R$, is calculated by using Eqs.1,3 (see Appendix 2):

$$E_{out} = \frac{\varepsilon \cdot \varepsilon_0}{2} \int_R^\infty \left[\frac{dV(Z,Q)}{dZ}\right]^2 4\pi Z^2 dZ =$$

$$\frac{k_e Q^2 \lambda_D}{\varepsilon R^2} e^{-2R/\lambda_D} \sinh^2\left(\frac{R}{\lambda_D}\right) \left[\frac{\lambda_D}{2R} + \frac{1}{4}\right] \quad (7)$$

## 3. Results

Here we calculate two types of electric energies: 1) $E_{CC}$, the interaction energy between the charges of the charged sphere and 2) $E_{in}$ and $E_{out}$, the electrolyte's field energy inside and around the charged sphere, respectively. We assume that the electrolyte contains only monovalent ions and calculate the electric energies in the following range of monovalent ion concentrations: $0.0001 - 100.0 [mol \cdot m^{-3}]$. Note, that the relative static permittivity of the electrolyte decreases with increasing ion concentration. However in the above concentration region the decrease is within one percent [8,9]. Thus in our calculations the relative static permittivity is taken as constant ($\varepsilon = 78$) at the above electrolyte concentrations. In this case, i.e. in the case of monovalent ions, the Debye length in $[m]$ is [10]:

$$\lambda_D = \left(\frac{\varepsilon_0 \varepsilon k_B T}{e^2 N_a 2C}\right)^{\frac{1}{2}} \quad (8)$$

where $\varepsilon_0 = 8.85 \cdot 10^{-12} [C^2 J^{-1} m^{-1}]$ is the vacuum permittivity, $\varepsilon$ is the relative static permittivity of the electrolyte, $k_B = 1.38 \cdot 10^{-23} [J K^{-1}]$ is the Boltzmann constant , $T[K]$ is the absolute temperature, $e = 1.6 \cdot 10^{-19} [C]$ is the charge of a positive monovalent ion, $N_a = 6 \cdot 10^{23} [mol^{-1}]$ is the Avogadro's number, $C [mol \cdot m^{-3}]$ is the monovalent ion concentration of the electrolyte inside and around the charged sphere. In our calculation we take always $T = 300 [K]$.

Based on Eq.5 in Figure 1 the charge-charge interaction energy is calculated as a function of the electrolyte concentration at three different radii of the charged sphere. In the case of our calculations the surface charge density of the charged sphere at every radius is $\rho_s = -0.266 \, [C \cdot m^{-2}]$. This is the charge density of PLFE (bipolar tetraether lipid with the polar lipid fraction E) vesicles if the cross sectional area of a PLFE is $0.6 nm^2$ and the charge of a PLFE molecule is $-1.6 \cdot 10^{-19} [C]$ (see refs.1,2,6).

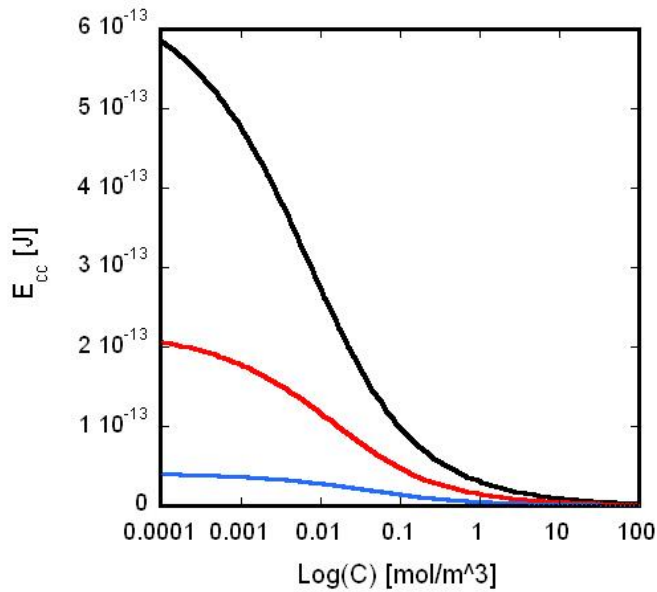

**Figure 1.** *Calculated charge-charge interaction energy $(E_{CC})$ of a charged sphere surrounded by electrolyte.* The energy is plotted against the ion concentration of the electrolyte. The radius of the charged sphere is: $R = 100 nm$ (black curve), $R = 70 nm$ (red curve), $R = 40 nm$ (blue curve). In the case of every charged sphere the charge density is the same: $-0.266 \, C/m^2$.

The field energy of the electrolyte inside and around the charged sphere is calculated by Eq.6 and Eq.7, respectively. These calculated energies are plotted against the ion concentration of the electrolyte (see Figures 2B and 2A).

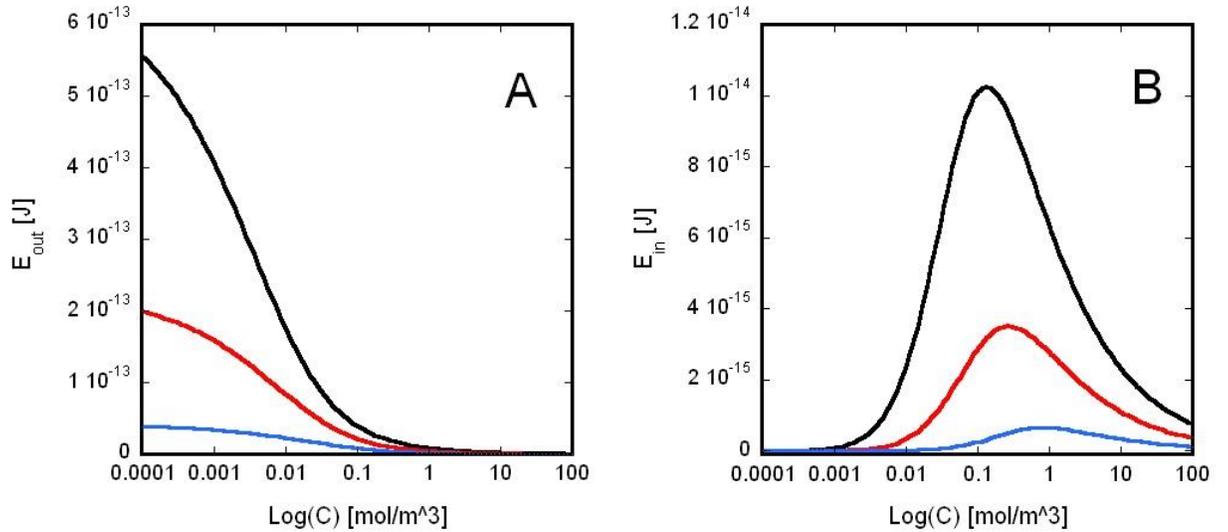

**Figure 2.** *Calculated field energy of the electrolyte around and inside the charged sphere.* The field energies of the electrolyte A) around ($E_{out}$) and B) inside ($E_{in}$) the charged sphere are plotted against the ion concentration of the electrolyte. The radius of the charged sphere is: $R = 100nm$ (black curve), $R = 70nm$ (red curve), $R = 40nm$ (blue curve). In the case of every charged sphere the charge density is the same: $-0.266 \, C/m^2$.

## 3    Discussion

By using Eqs.5,6 and 7 one can calculate the energy needed to build a charged sphere and the field energy of the electrolyte surrounding the sphere inside and around. In the case of zero ion concentration these energies can be obtained from the above mentioned equations by taking infinite long Debye length:

$$E_{cc} = \frac{k_e \cdot Q^2}{2 \cdot \varepsilon \cdot R^2} \cdot \lim_{\lambda_D \to \infty} \lambda_D \cdot e^{-\frac{R}{\lambda_D}} \cdot sinh\left(\frac{R}{\lambda_D}\right) =$$

$$\frac{k_e \cdot Q^2}{2 \cdot \varepsilon \cdot R^2} \cdot \lim_{\lambda_D \to \infty} \lambda_D \cdot e^{-\frac{R}{\lambda_D}} \cdot \left[\frac{R}{\lambda_D} + \frac{1}{3!}\left(\frac{R}{\lambda_D}\right)^3 + \frac{1}{5!}\left(\frac{R}{\lambda_D}\right)^5 + \cdots\right] =$$

$$\frac{k_e \cdot Q^2}{2 \cdot \varepsilon \cdot R} \tag{9}$$

$$E_{in} = \frac{k_e Q^2}{2\varepsilon R^2} \lim_{\lambda_D \to \infty}\left[e^{-\frac{2R}{\lambda_D}}\left\{\frac{\lambda_D}{2}\sinh\left(\frac{R}{\lambda_D}\right)\cosh\left(\frac{R}{\lambda_D}\right) + \frac{R}{2} - \frac{\lambda_D^2}{R}\sinh^2\left(\frac{R}{\lambda_D}\right)\right\}\right] =$$

$$\frac{k_e Q^2}{2\varepsilon R^2}\left\{\lim_{\lambda_D \to \infty} e^{-\frac{2R}{\lambda_D}} \cdot \cosh\left(\frac{R}{\lambda_D}\right)\cdot \frac{\lambda_D}{2}\cdot\left[\frac{R}{\lambda_D} + \frac{1}{3!}\left(\frac{R}{\lambda_D}\right)^3 + \frac{1}{5!}\left(\frac{R}{\lambda_D}\right)^5 + \cdots\right]\right\} +$$

$$\frac{k_e Q^2}{2\varepsilon R^2}\left[\frac{R}{2} - \lim_{\lambda_D \to \infty} e^{-\frac{R}{\lambda_D}}\frac{\lambda_D^2}{R}\left\{\frac{R^2}{\lambda_D^2} + \frac{2R}{3!\lambda_D}\left(\frac{R}{\lambda_D}\right)^3 + \cdots\right\}\right] =$$

$$\frac{k_e Q^2}{2\varepsilon R^2}\left[\frac{R}{2} + \frac{R}{2} - R\right] = 0 \tag{10}$$

$$E_{out} = \frac{k_e Q^2}{\varepsilon R^2}\lim_{\lambda_D \to \infty}\left\{e^{-2R/\lambda_D}\sinh^2\left(\frac{R}{\lambda_D}\right)\left[\frac{\lambda_D^2}{2R} + \frac{\lambda_D}{4}\right]\right\} =$$

$$\frac{k_e Q^2}{\varepsilon R^2}\lim_{\lambda_D \to \infty}\left\{e^{-2R/\lambda_D}\cdot\left\{\frac{R^2}{\lambda_D^2} + \frac{2R}{3!\lambda_D}\left(\frac{R}{\lambda_D}\right)^3 + \cdots\right\}\cdot\left[\frac{\lambda_D^2}{2R} + \frac{\lambda_D}{4}\right]\right\} =$$

$$\frac{k_e \cdot Q^2}{2 \cdot \varepsilon \cdot R} \tag{11}$$

The results in Eqs.10,11 are in accordance with the Shell Theorem [5]. According to the Shell Theorem at zero electrolyte ion concentration the electric potential inside the charged sphere is constant (see Eq.1 at ref.6). Thus the electric field strength is zero and according to Eq.3 the electric field energy is zero too. Also, according to the Shell Theorem at zero electrolyte ion concentration the electric potential around the charged sphere is $V(Z > R) = k_e Q/(\varepsilon Z)$ (see Eq.2 in ref.6). After substituting this into Eq.3 we get a result similar to Eq.11.

With increasing electrolyte ion concentration the electrical screening increases while the Debye length is approaching zero. As a consequence the energies, $E_{cc}, E_{in}$ and $E_{out}$, approach zero too (see Appendix 3).

In Figure 3 we compare the charge-charge interaction energy, $E_{cc}$ with the electrolyte's field energy within and outside the sphere (i.e.: $E_{in}/E_{cc}$ (dashed lines) and $E_{out}/E_{cc}$ (solid lines)) and with the polarization energy $E_{pol}/E_{cc}$ (dotted lines, see explanation below).

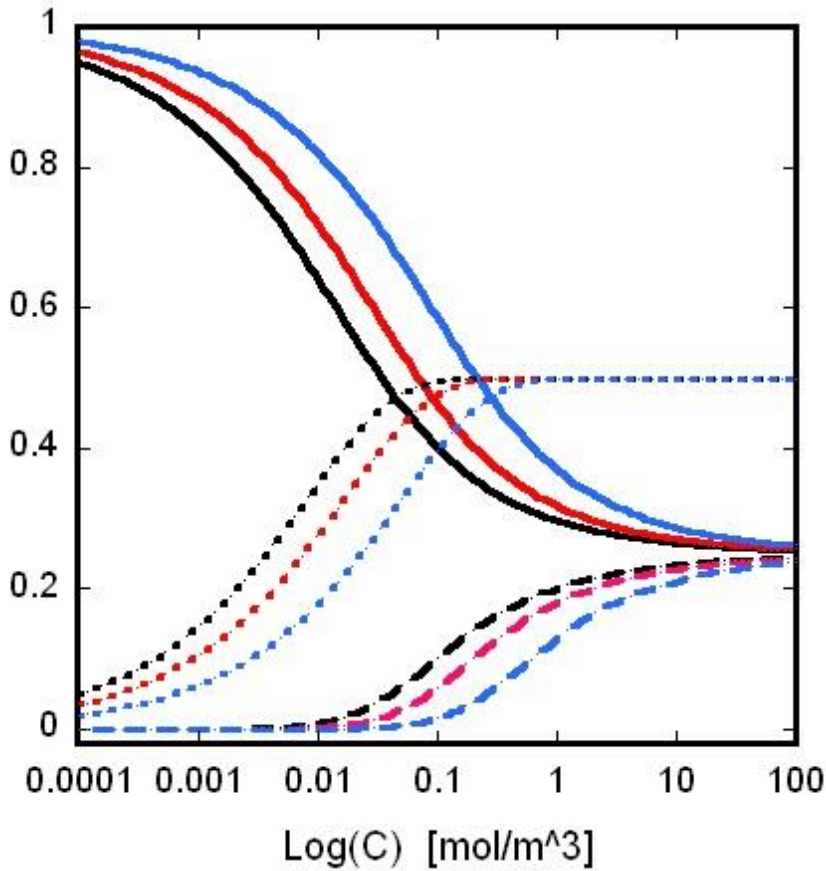

**Figure 3** *Electrolyte's field energies and polarization energy ($E_{in}$, $E_{out}$ and $E_{pol}$) relative to the charge-charge interaction energy ($E_{cc}$) are plotted against the ion concentration of the electrolyte (C).* Radius of the charged sphere: $100nm$ (black lines), $70nm$ (red line), $40nm$ (blue line). Solid lines: $E_{out}/E_{cc}$ vs. $C$. Dashed lines: $E_{in}/E_{cc}$ vs. $C$. Dotted lines: $E_{pol}/E_{cc}$ vs. $C$ (calculated by Eq.12).

According to Figure 3 close to zero ion concentration in the electrolyte $E_{cc} \cong E_{out}$ while $E_{in} \cong 0$, i.e. the energy needed to build up the charged sphere is close to the electrolyte's field energy around the sphere. In this case building up the charged sphere one works only against the repulsion of the charges already present in the sphere and the respective electrolyte's field energy appears only around the sphere.

However when the ion concentration of the electrolyte is higher than zero the work is expended also on the polarization of the electrolyte, $E_{pol}$. Namely, when the surface charge of the sphere is zero the average charge density in the electrolyte is zero everywhere. At negative surface charge of the sphere the average charge density close to the surface of the sphere is positive and it is decreasing with increasing distance from the surface. In this case the energy available to build up the electrolyte's field energy inside and around the charged sphere is: $E_{cc} - E_{pol}(= E_{in} + E_{out})$. Thus:

$$\frac{E_{pol}}{E_{CC}} = 1 - \left[\frac{E_{in}}{E_{CC}} + \frac{E_{out}}{E_{CC}}\right] \tag{12}$$

With increasing electrolyte ion concentration the Debye length decreases (see Eq.8) while the screening effect of the electrolyte increases and thus the work to build up the charged sphere, $E_{cc}$ decreases considerably (see Figure 1). From about $10 mol/m^3$ electrolyte ion concentration close to one quarter of the energy for building up the charged sphere appears as the electrolyte's field energy inside the sphere, while another quarter appears as the electrolyte's field energy around the sphere, and half of it appears as the polarization energy (see Figure 3).

The electrolyte's field energy $E_{out}$ also decreases with increasing ion concentration (see Figure 2A) because of the screening increases. But as it was mentioned above $E_{out}$ decreases more than $E_{cc}$ with increasing ion concentration (see Figure 3, solid lines).

Figure 2B shows the electrolyte's field energy inside the charged sphere as a function of the ion concentration of the electrolyte. Here the electrolyte's field energy is zero at zero ion concentration, $E_{in} = 0$ (Eq.10), because in this case the electric potential is constant inside the charged sphere (Eq.1 in ref.6) and thus the respective field strength is zero. At higher than zero ion concentration of the electrolyte $E_{in} > 0$ because the absolute value of the electric potential decreases toward the center of the sphere (Fig.3A in ref.6) and thus the respective field strength is not zero. However, at very high ion concentration $E_{in}$ reduces to close to zero. This is the case because of the increased screening the electric potential inside the sphere becomes close to zero.

Finally, Figures 1,2 show that the energies ($E_{cc}, E_{in}, E_{out}$) are decreasing with decreasing radius of the charged sphere while the charge density of the sphere, $\rho_e$ remains the same. This is the case because according to Eqs.5-7 these energies are proportional to the square of the total charge of the sphere. In the case of zero electrolyte ion concentration after substituting $Q = \rho_e 4\pi R^2$ into Eqs.9,11 we get $E_{cc} = E_{out} \sim R^3$.

## Conclusions

In this paper we considered a charged sphere surrounded by electrolyte. By using the Generalized Shell Theorem [6] analytical equations have been derived to calculate the electric energies of this system (Eqs.5-7). The work needed to build up the charged sphere, $E_{CC}$ decreases with increasing ion concentration of the electrolyte because the increasing screening effect on the charge-charge interaction. The energy needed to build up the charged sphere appears as sum of the electrolyte's field energy and the polarization energy of the ion's. At zero ion concentration the electrolyte's field energy around the charged sphere ($E_{out}$) is equal with $E_{CC}$, while the polarization energy ($E_{pol}$) and the electrolyte's field energy inside the sphere ($E_{in}$) are both equal with zero. At high ion concentrations ($C > 10[mol \cdot m^{-3}]$) half of the charge-charge interaction energy is for the polarization of ions and the other half is for the field energy of the electrolyte (i.e. $E_{out} \cong E_{in} \cong 0.25 \cdot E_{CC}$).

## Acknowledgements

The author is very thankful for Chinmoy Kumar Ghose

## References

[1] Gabriel JL, Chong PLG (2000) Molecular modeling of archaebacterial bipolar tetraether lipid membranes. *Chemistry and Physics of Lipids* 105: 193-200.

[2] Chong P (2010) Archaebacterial bipolar tetraether lipids: Physico-chemical and membrane properties. *Chem Phys Lipids* 163: 253–265.

[3] Almeida PFF (2009) Thermodynamics of lipid interactions in complex bilayers. *Biochim. Biophys. Acta* 1788: 72–85.

[4] Sugar IP, Thompson TE, Biltonen RL (1999) Monte Carlo simulation of two-component bilayers: DMPC/DSPC mixtures. *Biophysical Journal* 76: 2099-2110.

[5] Newton I (1999) A New Translation. *The Principia: Mathematical Principles of Natural Philosophy,* Berkeley: University of California Press, 590.

[6] Sugar IP. (2020) A generalization of the Shell Theorem. Electric potential of charged spheres and charged vesicles surrounded by electrolyte. *AIMS Biophysics* 7: 76-89.

[7] Griffiths DJ. (2007) *Introduction to Electrodynamics* (3rd Edition), Pearson Education, Dorling Kindersley

[8] Hasted JB, Ritson DM, Collie CH. (1948) Dielectric properties of aqueous ionic solutions. J. Chem. Phys. 16: 1-21.

[9] Gavish N, Promislow K. (2016), Dependence of the dielectric constant of electrolyte solutions on ionic concentration - a microfield approach. *arXiv:1208.5169v2 [physics.chem-ph] 1 Jul 2016.*

[10] Shohet JL. (2003), Encyclopedia of Physical Science and Technology (Third Edition)

[11] Moll VH (2014), *Special Integrals of Gradsteyn and Ryzhik: the Proofs* – Volume I and II. Series: *Monographs and Research Notes in Mathematics* (First Edition), Chapman and Hall/CRC Press

## *Appendix 1*

In order to calculate $E_{in}$ let us substitute the derivative of Eq.2 into Eq.3:

$$E_{in} = \frac{k_e Q^2 \lambda_D^2}{2\varepsilon R^2} e^{-2R/\lambda_D} \int_0^R \left[ -\frac{\sinh\left(\frac{Z}{\lambda_D}\right)}{Z^2} + \frac{\cosh\left(\frac{Z}{\lambda_D}\right)}{Z\lambda_D} \right]^2 Z^2 dZ \qquad (A1)$$

The integral in Eq.A1 can be separated to three terms:

$$\int_0^R \left[ -\frac{\sinh\left(\frac{Z}{\lambda_D}\right)}{Z^2} + \frac{\cosh\left(\frac{Z}{\lambda_D}\right)}{Z\lambda_D} \right]^2 Z^2 dZ = \int_0^R \frac{\sinh^2\left(\frac{Z}{\lambda_D}\right)}{Z^2} dZ - \int_0^R \frac{\sinh\left(\frac{2Z}{\lambda_D}\right)}{Z\lambda_D} dZ +$$

$$\int_0^R \frac{\sinh^2\left(\frac{Z}{\lambda_D}\right) + 1}{\lambda_D^2} dZ \qquad (A2)$$

The first term in Eq.A2 is:

$$\int_0^R \frac{e^{2Z/\lambda_D} - 2 + e^{-2Z/\lambda_D}}{4Z^2} dZ = \left[ -\frac{e^{2Z/\lambda_D}}{4Z} \right]_0^R + \frac{1}{2\lambda_D} \int_0^R \frac{e^{2Z/\lambda_D}}{Z} dZ + \left[ \frac{1}{2Z} \right]_0^R +$$

$$\left[-\frac{e^{-2Z/\lambda_D}}{4Z}\right]_0^R - \frac{1}{2\lambda_D}\int_0^R \frac{e^{-\frac{2Z}{\lambda_D}}}{Z}dZ = \frac{1}{2\lambda_D}\int_0^R \frac{e^{\frac{2Z}{\lambda_D}}}{Z}dZ - \frac{1}{2\lambda_D}\int_0^R \frac{e^{-\frac{2Z}{\lambda_D}}}{Z}dZ +$$

$$\left[\frac{1-e^{\frac{2Z}{\lambda_D}}}{4Z}\right]_0^R + \left[\frac{1-e^{-\frac{2Z}{\lambda_D}}}{4Z}\right]_0^R =$$

$$\frac{1}{2\lambda_D}\int_0^R \frac{e^{\frac{2Z}{\lambda_D}}}{Z}dZ - \frac{1}{2\lambda_D}\int_0^R \frac{e^{-\frac{2Z}{\lambda_D}}}{Z}dZ + \frac{1}{2R}\left[1-\cosh\left(\frac{2R}{\lambda_D}\right)\right] \tag{A3}$$

Note that above we used [11]: $\int \frac{e^{ax}}{x^2}dx = \left(-\frac{e^{ax}}{x} + a\int \frac{e^{ax}}{x}dx\right)$.

Calculating the last term in Eq.A3 we used the following two limits:

$$\lim_{Z\to 0}\left[\frac{1-e^{\frac{2Z}{\lambda_D}}}{Z}\right] = \lim_{Z\to 0}\frac{1-\left[1+\frac{1}{1!}\left(\frac{2Z}{\lambda_D}\right)^1+\frac{1}{2!}\left(\frac{2Z}{\lambda_D}\right)^2+\frac{1}{3!}\left(\frac{2Z}{\lambda_D}\right)^3+\cdots\right]}{Z} = -\frac{2}{\lambda_D} \tag{A4}$$

and

$$\lim_{Z\to 0}\left[\frac{1-e^{-\frac{2Z}{\lambda_D}}}{Z}\right] = \lim_{Z\to 0}\frac{1-\left[1-\frac{1}{1!}\left(\frac{2Z}{\lambda_D}\right)^1+\frac{1}{2!}\left(\frac{2Z}{\lambda_D}\right)^2-\frac{1}{3!}\left(\frac{2Z}{\lambda_D}\right)^3+\cdots\right]}{Z} = \frac{2}{\lambda_D} \tag{A5}$$

The second term in Eq.A2 is:

$$-\int_0^R \frac{\sinh\left(\frac{2Z}{\lambda_D}\right)}{Z\lambda_D}dZ = \frac{1}{2\lambda_D}\int_0^R \frac{e^{-\frac{2Z}{\lambda_D}}}{Z}dZ - \frac{1}{2\lambda_D}\int_0^R \frac{e^{\frac{2Z}{\lambda_D}}}{Z}dZ \tag{A6}$$

The third term in Eq.A2 is:

$$\int_0^R \frac{\sinh^2\left(\frac{Z}{\lambda_D}\right)}{\lambda_D^2}dZ + \int_0^R \frac{1}{\lambda_D^2}dZ = \frac{1}{\lambda_D^2}\left[\frac{\lambda_D}{2}\sinh\left(\frac{Z}{\lambda_D}\right)\cosh\left(\frac{Z}{\lambda_D}\right) + \frac{Z}{2}\right]_0^R \tag{A7}$$

After summarizing the three terms (Eqs.A3,A6,A7) of the integral in Eq.A1 we get $E_{in}$:

$$E_{in} = \frac{k_e Q^2 \lambda_D^2}{2\varepsilon R^2} e^{-\frac{2R}{\lambda_D}} \left\{ \frac{1}{2\lambda_D} \sinh\left(\frac{R}{\lambda_D}\right) \cosh\left(\frac{R}{\lambda_D}\right) + \frac{R}{2\lambda_D^2} + \frac{1}{2R}\left[1 - \cosh\left(\frac{2R}{\lambda_D}\right)\right] \right\} =$$

$$\frac{k_e Q^2 \lambda_D^2}{2\varepsilon R^2} e^{-\frac{2R}{\lambda_D}} \left\{ \frac{1}{2\lambda_D} \sinh\left(\frac{R}{\lambda_D}\right) \cosh\left(\frac{R}{\lambda_D}\right) + \frac{R}{2\lambda_D^2} - \frac{1}{R}\sinh^2\left(\frac{R}{\lambda_D}\right) \right\} \qquad (A8)$$

## Appendix 2

In order to calculate $E_{out}$ let us substitute the derivative of Eq.1 into Eq.3:

$$E_{out} = \frac{k_e Q^2 \lambda_D^2}{2\varepsilon R^2} \sinh^2\left(\frac{R}{\lambda_D}\right) \int_R^\infty e^{-2Z/\lambda_D} \left[\frac{1}{Z} + \frac{1}{\lambda_D}\right]^2 dZ =$$

$$\frac{2k_e Q^2}{\varepsilon R^2} \sinh^2\left(\frac{R}{\lambda_D}\right) \int_R^\infty e^{-\frac{2Z}{\lambda_D}} \left[\frac{\lambda_D^2}{4Z^2} + \frac{\lambda_D}{2Z} + \frac{1}{4}\right] dZ \qquad (A9)$$

After substituting $2Z/\lambda_D$ by $\omega$ in Eq.A9 we get:

$$E_{out} = \frac{\lambda_D k_e Q^2}{\varepsilon R^2} \sinh^2\left(\frac{R}{\lambda_D}\right) \left\{ \int_{\frac{2R}{\lambda_D}}^\infty \frac{e^{-\omega}}{\omega^2} d\omega + \int_{\frac{2R}{\lambda_D}}^\infty \frac{e^{-\omega}}{\omega} d\omega + \int_{\frac{2R}{\lambda_D}}^\infty \frac{e^{-\omega}}{4} d\omega \right\} =$$

$$\frac{\lambda_D k_e Q^2}{\varepsilon R^2} \sinh^2\left(\frac{R}{\lambda_D}\right) \left\{ \left[-\frac{e^{-\omega}}{\omega}\right]_{\frac{2R}{\lambda_D}}^\infty - \int_{\frac{2R}{\lambda_D}}^\infty \frac{e^{-\omega}}{\omega} d\omega + \int_{\frac{2R}{\lambda_D}}^\infty \frac{e^{-\omega}}{\omega} d\omega + \int_{\frac{2R}{\lambda_D}}^\infty \frac{e^{-\omega}}{4} d\omega \right\} =$$

$$\frac{\lambda_D k_e Q^2}{\varepsilon R^2} \sinh^2\left(\frac{R}{\lambda_D}\right) e^{-\frac{2R}{\lambda_D}} \left[\frac{\lambda_D}{2R} + \frac{1}{4}\right] \qquad (A10)$$

Note that above we used [11]: $\int \frac{e^{ax}}{x^2} dx = \left(-\frac{e^{ax}}{x} + a\int \frac{e^{ax}}{x} dx\right)$.

## Appendix 3

Here we calculate the energies, $E_{cc}, E_{in}$ and $E_{out}$, when $\lambda_D$ approaches zero.

$$E_{cc} = \frac{k_e \cdot Q^2}{2 \cdot \varepsilon \cdot R^2} \cdot \lim_{\lambda_D \to 0} \lambda_D \cdot e^{-\frac{R}{\lambda_D}} \cdot \sinh\left(\frac{R}{\lambda_D}\right) =$$

$$\frac{k_e \cdot Q^2}{2 \cdot \varepsilon \cdot R^2} \cdot \lim_{\lambda_D \to 0} \lambda_D \cdot \frac{1 - e^{-\frac{2 \cdot R}{\lambda_D}}}{2} = 0 \qquad (A11)$$

$$E_{in} = \frac{k_e Q^2}{2\varepsilon R^2} \lim_{\lambda_D \to 0}\left[ e^{-\frac{2R}{\lambda_D}} \left\{ \frac{\lambda_D}{2} \sinh\left(\frac{R}{\lambda_D}\right) \cosh\left(\frac{R}{\lambda_D}\right) + \frac{R}{2} - \frac{\lambda_D^2}{R} \sinh^2\left(\frac{R}{\lambda_D}\right) \right\} \right] =$$

$$\frac{k_e Q^2}{2\varepsilon R^2} \lim_{\lambda_D \to 0} e^{-\frac{2R}{\lambda_D}} \left[ \frac{\lambda_D}{2} \cdot \frac{e^{\frac{2 \cdot R}{\lambda_D}} - e^{-\frac{2 \cdot R}{\lambda_D}}}{4} + \frac{R}{2} - \frac{\lambda_D^2}{R} \cdot \frac{e^{\frac{2 \cdot R}{\lambda_D}} - 2 + e^{-\frac{2 \cdot R}{\lambda_D}}}{4} \right] =$$

$$\frac{k_e Q^2}{2\varepsilon R^2} \lim_{\lambda_D \to 0} \left[ \frac{\lambda_D}{2} \cdot \frac{1 - e^{-\frac{4 \cdot R}{\lambda_D}}}{4} + \frac{R}{2} \cdot e^{-\frac{2R}{\lambda_D}} - \frac{\lambda_D^2}{R} \cdot \frac{1 - 2 \cdot e^{-\frac{2R}{\lambda_D}} + e^{-\frac{4 \cdot R}{\lambda_D}}}{4} \right] = 0 \quad (A12)$$

$$E_{out} = \frac{k_e Q^2}{\varepsilon R^2} \lim_{\lambda_D \to 0} \left\{ \sinh^2\left(\frac{R}{\lambda_D}\right) \left[ \frac{\lambda_D^2}{2R} + \frac{\lambda_D}{4} \right] \right\} e^{-2R/\lambda_D} =$$

$$\frac{k_e Q^2}{\varepsilon R^2} \lim_{\lambda_D \to 0} \left\{ \frac{1 - 2 \cdot e^{-\frac{2R}{\lambda_D}} + e^{-\frac{4 \cdot R}{\lambda_D}}}{4} \cdot \left[ \frac{\lambda_D^2}{2R} + \frac{\lambda_D}{4} \right] \right\} = 0 \qquad (A13)$$